# Interfaces in epitaxially grown $Zn_3P_2$ nanowires and their composition dependent optoelectronic properties for photovoltaic applications

Simon Escobar Steinvall[1,2], Francesco Salutari[3], Jonas Johansson[2,4], Ishika Das[5], Sebastian Lehmann[2,4], Stephen A. Church[5], M. Chiara Spadaro[3,6], Patrick Parkinson[5], Jordi Arbiol[3,7], Kimberly A. Dick[1,2]

1. Center for Analysis and Synthesis, Lund University, Box 124, 221 00 Lund, Sweden
2. NanoLund, Lund University, Lund, Sweden
3. Catalan Institute of Nanoscience and Nanotechnology (ICN2), CSIC and BIST, Campus UAB, Bellaterra, 08193 Barcelona, Catalonia, Spain
4. Division of Solid State Physics, Lund University, 221 00 Lund, Sweden
5. Department of Physics and Astronomy and The Photon Science Institute, The University of Manchester, Manchester, M13 9PL United Kingdom
6. Department of Physics and Astronomy "Ettore Majorana", University of Catania and CNR-IMM Via S. Sofia 64, 95123 Catania, Italy
7. ICREA, Pg. Lluís Companys 23, 08010 Barcelona, Catalonia, Spain

**Abstract**

Epitaxially grown nanowires have shown promise for photovoltaic applications due to their nanophotonic properties. Moreover, the mechanical properties of nanowires can reduce crystallographic defect formation at interfaces to help enable new material combinations for photovoltaics. One material that stands to benefit from the nanowire morphology is zinc phosphide ($Zn_3P_2$), which despite promising optoelectronic properties has experienced limited applicability due to challenges achieving heteroepitaxy, stemming from its large lattice parameter and coefficient of thermal expansion. Herein, we identify the requirements for successful epitaxy of $Zn_3P_2$ nanowires using metalorganic chemical vapour deposition and the impact on interface structure and defect formation. Furthermore, using high-throughput optical spectroscopy we were able to demonstrate shifts in the photoluminescence intensity and energy by tuning the V/II ratio during growth, highlighting the compositional tunability of the optoelectronic properties of $Zn_3P_2$ nanowires.

## 1. Introduction

The application of photovoltaics is an integral part in more sustainable electricity generation. The cost of installing new solar panels is continuously decreasing, and coupled with the increased demand, the roll-out of new photovoltaic modules is increasing exponentially.[1] This development is dominated by Si-based modules, which take advantage of the maturity of the material and the established supply routes. However, the increasing demand is pushing existing polysilicon supply routes to their limit.[2] Moreover, the indirect bandgap of Si complicates its applicability in flexible photovoltaics and other light-weight applications without the use of additional light-trapping mechanisms.[3] Therefore, to diversify and expand the applicability of the photovoltaic sector there is currently significant research done into alternative earth-abundant semiconductors with direct band gaps, one potential candidate being zinc phosphide ($Zn_3P_2$).[4,5]

$Zn_3P_2$ has optoelectronic properties suitable for single-junction photovoltaics, such as a direct band gap of 1.5 eV[6,7], high optical absorptance[8] and long carrier diffusion lengths[9]. However, charge extraction is still challenging due to i) its intrinsic p-doping making n-doping and p-n junctions difficult to achieve[10,11] and ii) its large lattice parameter and coefficient of thermal expansion (CTE), which complicates $Zn_3P_2$'s incorporation in heterostructures without defect formation.[12,13] Recently, there have been investigations that utilise various nanoscale epitaxy approaches and molecular beam epitaxy to overcome the limitations set by the lattice parameter and CTE.[14–19] One such approach, vapour-liquid-solid (VLS) growth of epitaxial $Zn_3P_2$ nanowires, which was successfully implemented to grow vertical single-crystal nanowires with tuneable composition and optoelectronic properties. However, previous reports on this approach reported limited control over the growth directions and nanowire density, which is necessary to control to take advantage of their nanophotonic properties and integrate them into devices.[14,20–22] To properly evaluate $Zn_3P_2$ nanowires potential in actual devices we therefore need to gain increased control over these parameters, and also transfer the synthesis to more large-scale and high throughput techniques, such as metalorganic chemical vapour deposition (MOCVD).

In this report we explored the necessary steps for $Zn_3P_2$ nanowire growth in MOCVD, and mapped the parameter space through a combinatorial study, focusing on temperature and partial pressures of the precursors, supported by thermodynamic modelling. Furthermore, electron microscopy is used for structural characterisation and high-throughput optical spectroscopy for functional characterisation to evaluate the quality and suitability of the grown material in photovoltaic applications.

**2. Experimental**

The growth was performed on InP (111)A and B substrates (Wafer Technology Ltd.) using an Aixtron 3x2" close-coupled showerhead (CCS) MOCVD system operating at a pressure of 100 mbar, constant flow rate of 8000 sccm and an InP base coverage (the effect of the base cover is expanded on in the SI). First, the native oxide was removed through degassing the samples under a $PH_3$ partial pressure of 0.1 mbar for 7 minutes at 530 °C. Next, we deposited self-assembled In particles at 380 °C for 240 s and a TMIn partial pressure of $2.5 \times 10^{-3}$ mbar to act as the catalyst for nanowire growth. The temperature was subsequently lowered to the set growth temperature, which ranged from 305°C to 355 °C. Once the temperature had stabilised we performed a 5 minute Zn pre-deposition step at a DEZn partial pressure of $1.0 \times 10^{-2}$ mbar and closed $PH_3$ supply. The nanowire growth was then initiated by again supplying $PH_3$, with varying partial pressure depending on the experiment, while maintaining the aforementioned DEZn partial pressure. The $PH_3$ partial pressures were varied from $1.25 \times 10^{-3}$ to 0.5 mbar for growth times spanning 1 minute to 120 minutes.

Scanning electron microscopy (SEM) was performed using a Zeiss Gemini 500 SEM operating at 5 kV using an in-lens detector.

The phase diagrams were calculated with the Thermo-Calc software using the same approach as in ref. [14], that is, by extrapolating from binary phase data.

Electron transparent lamellae of longitudinal section of nanowires were obtained via focused ion beam (FIB) processing using a Helios UX 5 FIB system. A pre-deposition of 0.2 μm of carbon (nominal thickness) and 0.2 μm of tungsten(W) (nominal thickness) was performed with the electron gun inside the FIB system prior to the lamella processing. This procedure aimed at limiting the contamination from the thick ion-deposited W protective layer during the final stages of the thinning.

Aberration corrected HAADF-STEM images were obtained using a Thermo Fisher Scientific Spectra 300 double corrected transmission electron microscope (TEM). Denoised high magnification images were obtained by filtering the corresponding power spectra with spot masks and taking the inverse FFT. Electron Energy Loss Spectroscopy in STEM mode (EELS-STEM) compositional maps were obtained in a FEI Tecnai F20 TEM by using a GATAN QUANTUM filter. 3D atomic models were created using CaRIne Crystallography 3.1.[23] Geometrical phase analysis (GPA) was performed on simulated and experimental images by using the licensed GPA plug-in available for Gatan Digital Micrograph.[24]

The optoelectronic properties of the nanowires were assessed using room temperature photoluminescence (PL) spectroscopy in a confocal microscope setup, similar to that discussed in Ref. [25]. Photoexcitation was achieved using a 532 nm continuous wave laser, with an optical power of 3 mW, focussed to a diffraction limited spot with a 100x magnification objective lens. The luminescence was collected using an optical fibre and the spectrum was measured using a Horiba iHR550 spectrometer with a 150 lines/mm diffraction grating and a slit width of 1 mm. 30 μm by 30 μm micro-PL maps were measured by translating samples in steps of 0.5 μm in x and y.

**3. Results and discussion**

**3.1 Nanowire growth**

Once the substrate had been degassed under a $PH_3$ atmosphere in the MOCVD to remove the native oxide, the next step of the process is the deposition of the In-catalyst particles, as shown in Figure 1a. The self-assembled In particles had a density of 4.7 particles μm$^{-2}$ with a diameter distribution of 124 ± 66 nm on InP (111)B and a density of 3.9 particles μm$^{-2}$ with a diameter distribution of 138 ± 34 nm on InP (111)A for identical deposition conditions. In the subsequent step the temperature was lowered to the growth temperature. In initial trials we tried to start the growth by supplying both DEZn and PH$_3$ simultaneously from the start. However, this led to the results seen in the SEM image in Figure 1b and Figure S1, namely the formation of small nanopyramids. To promote nanowire growth (Figure 1c) we implemented a 5-minute Zn pre-deposition step. To understand the importance of this step we calculated the isothermal section of the Zn-In-P ternary phase diagram for our standard growth temperature (330 °C) shown in Figure 1d. To understand the initial nanowire growth mechanism we closely observe the corner closest to pure In (bottom left, zoomed plot in Figure 1e). We observe that if we start to provide Zn and P to the system we will enter a phase region where the stable phases are L + InP, meaning that we will start consuming our catalyst particle to precipitate InP, which explains the tapered shaped nanopyramids observed in Figure 1b. On the other hand, by first providing Zn through the pre-deposition we instead move into the top phase region in Figure 1e, which consists of L + Zn$_3$P$_2$, where Zn$_3$P$_2$ nanowire growth can be achieved, which is consistent with the SEM data shown in Figure 1c.

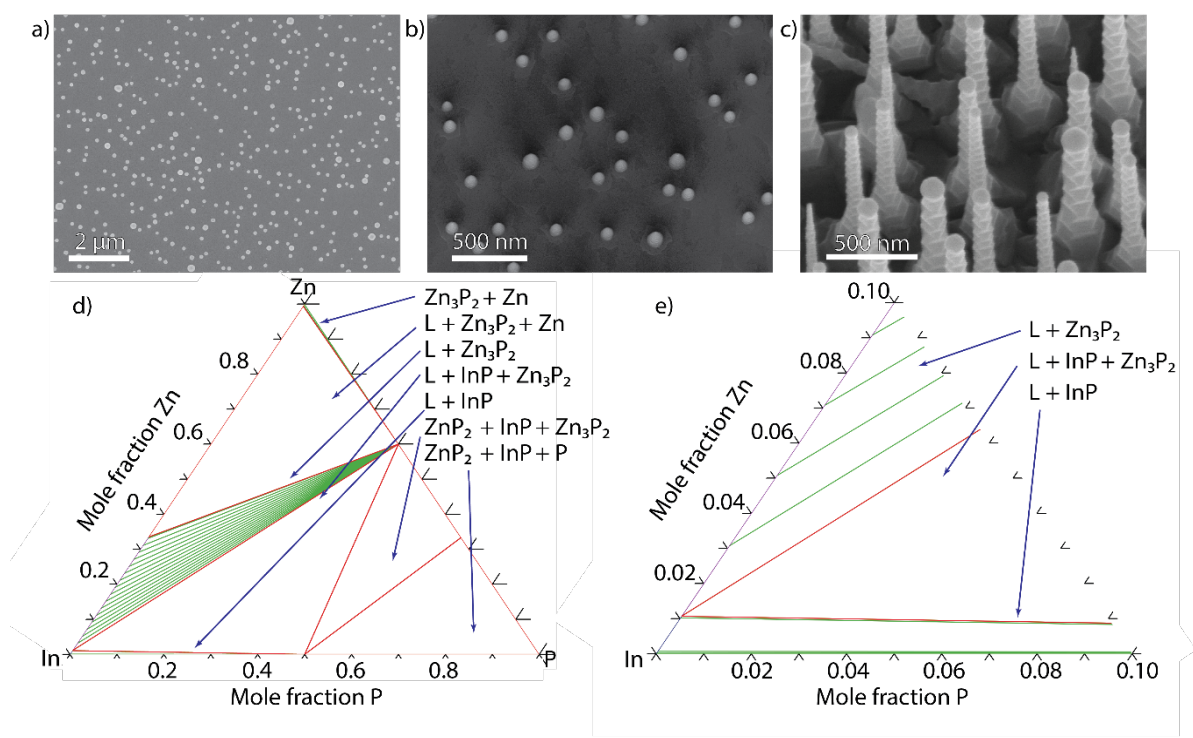

**Figure 1.** Scanning electron microscopy images of (a) as-deposited In nanoparticles on InP (111)B (top view), (b) pyramids grown without Zn pre-deposition (15° tilt) and (c) nanowires grown with Zn pre-deposition (30° tilt). (d) shows the calculated Zn-In-P ternary phase diagram at 330 °C and (e) shows a zoomed in section closest to pure In.

Another aspect worth noting is that we do not observe the co-existence of ZnP$_2$ and a liquid in the ternary phase diagram. This would explain why in In-catalysed Zn$_3$P$_2$ nanowires there have been no observation of the other stoichiometry irrespective of growth conditions in this or previous studies (although for different catalyst materials, such as Bi, it has been demonstrated).[14,26–28]

We explored the growth parameter space by tuning the V/II ratio and temperature at constant DEZn partial pressure (1.0×10$^{-2}$ mbar) and temperature (305-355 °C). First, we varied the PH$_3$ partial pressure from 1.25×10$^{-3}$ to 0.5 mbar, corresponding to nominal V/II ratios of 1.25 to 50.1 at a growth temperature of 330 °C. What we observed at lower PH$_3$ partial pressures was a low density of short nanowires with no obvious tapering of the diameter, as shown in Figure 2a. However, with increasing PH$_3$ we initially saw a significant increase of the growth rate, up to a PH$_3$ partial pressure of 0.25 mbar (V/II ratio of 25.1). For even higher PH$_3$ flows we still saw significant axial nanowire growth, but the tapering due to radial vapour-solid (VS) growth started to increase. We also start to observe the absence of the catalyst particle on some of the nanowires (see Figure 2a at V/II of 45.1 and Figure S1). This could possibly occur through consumption of the catalyst particle due to the high PH$_3$ partial pressure, and results in a stop in the axial growth partway through the run as shown in Figure S1.

When we instead explored the effect of temperature for a constant V/II ratio (PH$_3$ partial pressure of 0.35 mbar, V/II ratio of 35.1), we saw that increasing the temperature rapidly suppressed the growth of nanowires. As can be seen in Figure 2b, we observe the formation of small stub-like structures and we could not observe any catalyst particles. This is attributed to the increased desorption of the Zn atoms on the surface in combination with the rapid increase of the cracking efficiency of PH$_3$, leading to a much higher local V/II ratio at the growth interface. Consequently, the drastic change in local conditions will limit the VLS growth in favour of VS growth, resulting in thin film formation. The increased temperature and decreased Zn also has the potential to push the growth into phase regions which allow for the formation of InP (to the right corner of the phase diagrams shown in Figure 1d-e), which would consume the droplet and again limit the window for sustained VLS growth. On the other hand, by lowering the temperature we continued to observe nanowire growth in the explored range. However, the growth rate was decreasing significantly with decreasing temperature. When going from a growth temperature of 336 °C to 304 °C we observed a decrease of average nanowire length from 1050 ±62 nm to 800 ± 83 nm for a growth time of 30 minutes. This is most likely related to the strong decrease in the PH$_3$ cracking efficiency in this temperature range, which affects the material supply and growth rate.[29]

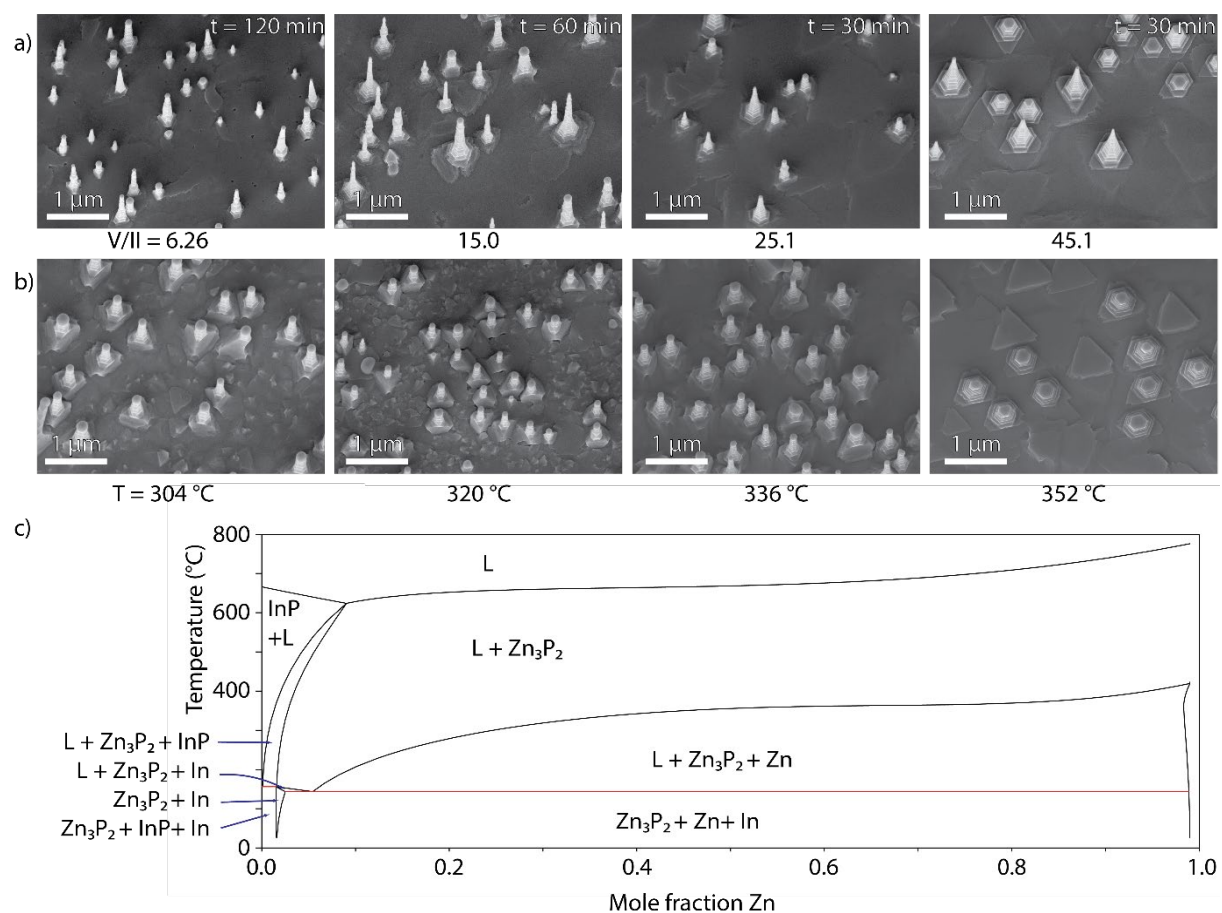

**Figure 2.** (a) 15° tilted SEM images of the V/II series at 330 °C and constant Zn flow. (b) 15° tilted SEM images of temperature series at constant V/II ratio of 35.1. (c) Plot of the calculated pseudo-binary Zn-In phase diagram at 1% P.

The temperatures and V/III ratios needed for MOCVD growth of $Zn_3P_2$ were higher than those reported in previous literature for growth using molecular beam epitaxy (MBE). We expect that the difference in growth temperature between the techniques is related to the decomposition of the $PH_3$, setting the lower limit of the parameter window for MOCVD. However, while there is decomposition of the precursors even at the lower end of the explored temperature range, the low efficiency results in the need for high $PH_3$ partial pressures to compensate for the lowered decomposition efficiency. This may contribute to the need for the higher nominal V/III ratios in MOCVD compared to MBE in the initial material supply. However, this will not necessarily result in a large difference in the local V/III ratio at the growth interface due to the low utilisation of the $PH_3$. To better understand the growth window, we can extract the pseudo-binary $Zn_3P_2$-In phase diagram with a constant P concentration of 1% to simulate the composition of the droplet during growth, presented in Figure 2c. In this phase diagram we can observe a growth window which contains a liquid phase and a $Zn_3P_2$ solid phase (L + $Zn_3P_{2(s)}$ region). For Zn concentrations in the order of a few percent in the droplet, the growth temperature can in theory go as low as 140 °C while the upper limit is set by the transition into the L + $InP_{(s)}$ + $Zn_3P_{2(s)}$ region where the In making up the droplet would be consumed to form InP. As seen in the temperature series in Figure 2b, we start crossing into this region before 350 °C, which indicates that we have approximately ~3-5% Zn in the droplet during growth. For higher Zn concentrations the growth window moves up in temperature to avoid the formation of Zn (lower limit) and because it is less favourable to form InP (upper limit). This does indicate that the theoretical growth window for $Zn_3P_2$ nanowires is very extensive, however, practical limitations with precursor supplies, mainly Zn re-evaporation and $PH_3$ cracking, limit the experimental parameter space by MOCVD.

Another big difference between the current and previous studies is that we could significantly suppress the appearance of multiple morphologies of epitaxial $Zn_3P_2$ nanowires in the same experiment as observed previously.[14] In previous reports, the catalyst particles were generated through the decomposition of the InP substrate.[14] However, in this study we developed a separate deposition step of the catalyst particle, which maintains a pristine interface that in turn favours only one type of growth. As schematically illustrated in Figure 3a, previous approaches could potentially expose multiple possible facets that could initiate nanowire growth, resulting in a mixed morphology distribution, similar to what was observed for InP nanowires growing on InP (001).[30] On the InP (111)A/B substrates used in this study we mainly observed the zigzag heterotwin-plane superlattice nanowires growing perpendicular to the substrate and the (101) plane of the $Zn_3P_2$.[31] In theory, we should have been able to access a twin-free morphology as described in Ref [14] by lowering the temperature. However, due to the lower temperature limit set by the $PH_3$ decomposition we were unable to observe them experimentally. Only at very low V/III ratios, such as those shown in Figure 2a, we could observe short segments without the zigzag morphology. We did however observe a new morphology of nanowires, shown in Figure 3b-c. These grew at an angle on the InP (111)A/B substrates. We believe this nanowire morphology is the result of a defect formation in the initial nucleus explained in more detail below.

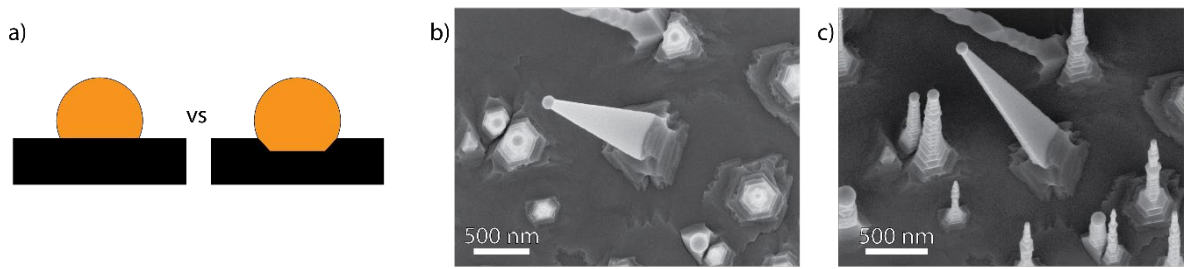

**Figure 3.** (a) Diagram depicting nucleation on flat vs mixed facet. SEM images of the new morphology showing (b) top view and (c) 15° tilted view.

### 3.2 Structural characterisation

We investigated the epitaxial relation between the $Zn_3P_2$ nanowires and the InP (111) A and B surfaces, as well as the defect structure using AC-HAADF-STEM. Images of the interfaces with InP (111) of different polarity and the base of a nanowire are presented in Figure 4a-b. We have observed that the $Zn_3P_2$ adapts with a good lattice match and clear epitaxial relation to the InP below. However, we have observed small differences depending on the polarity selected. In A-polar InP the surface is terminated with In, while in B-polar it is terminated with P. In the A-polar case, it seems there is an In-Zn interface, while in the case of the B-polar, the interface looks more like P-Zn (truncated unit cell). Moreover, we consistently observed a twin plane close to the interface between the InP (111)A and $Zn_3P_2$ nanowires, as highlighted in the higher magnification image in Figure 4c. However, this is not observed for growth on InP (111)B (Figure 4d) or in the thin film that grew in the regions between nanowires. In VLS growth of III-V and II-VI materials, one of the polarities has an increased probability to show twinning, especially in conditions out of steady-state (initial and final stages of nanowire growth).[32] In addition, the formation of twins predominantly happens in one of the polarities.[32] In GaAs it has been demonstrated that twinning predominantly occurs along (111)B planes, but rarely along (111)A.[33,34] In other systems, like GaSb[35] or InP[36,37], it is the opposite and twins more readily form along (111)A (without the presence of dopants)[36]. In the present case, we suspect that initial nucleation from the In-rich catalyst particle, when the growth process starts, may first produce a twinned InP layer before the $Zn_3P_2$ growth. In A-polar substrates, this first InP layer precipitates out of steady-state (initial growth stages), so, with high probability of twinning (as observed). In the B-polar case, this phenomenon would be rare as shown in the present experimental results.

To investigate the residual strain in the nanowire we performed geometric phase analysis (GPA) and 3D modelling on $Zn_3P_2$ grown on InP (111)B, compiled in Table 1. From the GPA we could not observe any misfit dislocations or other regular interface-related defects (Figure S2). However, we observe that the $Zn_3P_2$ close to the interface experiences slight strain. When the nanowires grow the $Zn_3P_2$ becomes progressively more relaxed, as explained by the radial strain relaxation observed in nanowires. The relaxation is observed for both nanowire morphologies. From the GPA analysis we also estimate the relative distortions in terms of dilation and rotation of the nanowires with respect to the substrate, as well as the magnitude of plane mismatch by considering the parallel and perpendicular planes with respect to the InP (111) surface, namely the (011) and (100) planes of $Zn_3P_2$. From this we can also calculate the strain tensor components $\varepsilon_{xx}$ and $\varepsilon_{yy}$, corresponding to the parallel and perpendicular directions with respect to the substrate, respectively. The residual values of the plane mismatch computed considering the (011) $Zn_3P_2$ plane indicate a minimal compression within the $Zn_3P_2$ layer. Along the (100) $Zn_3P_2$ plane, perpendicular to the interface, we instead obtained a positive value of the residual plane mismatch. The positive value indicates an overall expansion of the $Zn_3P_2$ layer in this direction. The rotation of $Zn_3P_2$ relative to the substrate is negligible for both the parallel and perpendicular directions.

We also performed the same analysis on the $Zn_3P_2$ grown on InP (111)A. However, the distorted interface under the nanowires due to the twins did not yield any useful quantitative results. Instead, the comparison with the area growing between nanowires turned out to be more interesting, shown in Figure 4e-f. In the absence of the twins and surface re-arrangement as observed in the nanowires we could instead clearly observe the formation of misfit dislocations along the interface between the thin film and the substrate. This would indicate that the re-arrangement of the interface under the nanowire and the stress relaxation are useful mechanisms in mitigating defect formation even at significantly greater diameters compared to those proposed previously.[17,38]

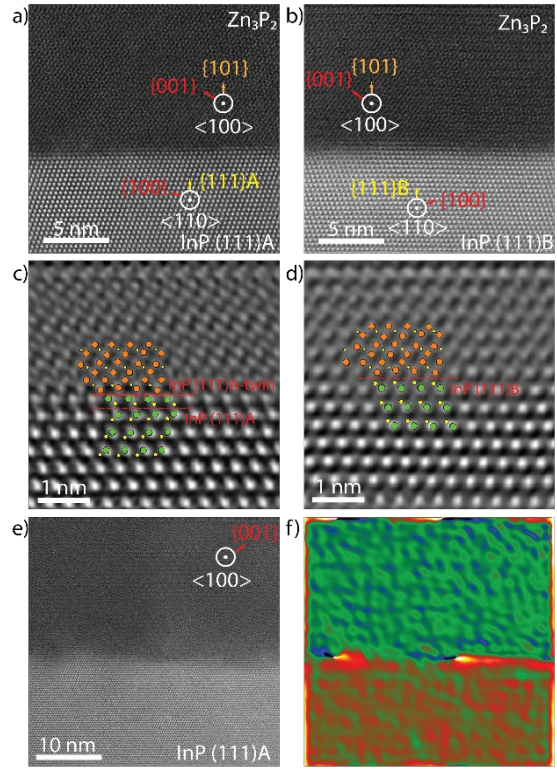

**Figure 4.** AC HAADF-STEM images of the interface between (a) InP (111)A or (b) InP (111)B and $Zn_3P_2$ viewed along a <110> zone axis with respect to the InP (<100> with respect to the $Zn_3P_2$). Filtered and high-magnification images of the interfaces of (c) InP (111)A and (d) InP (111)B and $Zn_3P_2$ nanowires highlighting the twin and interface atomic structure. (e) Atomic resolution AC HAADF STEM image of the InP (111)A substrate and thin film between nanowires and corresponding GPA map (f) highlighting the misfit dislocations.

**Table 1.** Summary of GPA results along different directions and the residual strain of $Zn_3P_2$ nanowires grown on InP (111)B.

|  | Plane mismatch | | $\varepsilon_{xx}$ | $\varepsilon_{yy}$ |
|---|---|---|---|---|
| **Direction** | $(011)_{Zn_3P_2}//(111)_{InP}$ | $(100)_{Zn_3P_2}//(1-10)_{InP}$ | | |
| **GPA Measured** | -3.0% | -2.3% | -2.3% | -2.9% |
| **Fully relaxed (theory)** | -2.9% | -2.7% | -2.6% | -2.9% |
| **Residual strain** | -0.1% | 0.4% | 0.3% | 0.0% |

We extended the HAADF-STEM analysis to the bulk of the tilted nanowires discussed at the end of section 3.1, shown in Figure 5. These nanowires grew with a tilt with respect to the InP (111) surface and without the twin plane superlattice structure. However, we observed multiple types of defects within these nanowires. First, we noted a twin with a similar structure to previously observed heterotwins, running along the growth axis of the nanowire, shown in Figures 5a-c.[31] We observed it reaching very close, but not all the way to the InP-$Zn_3P_2$ interface in our FIB cross sections. However, due to the nature of the fabrication process we removed a significant part of the nanowire in the milling step to achieve the electron transparent sample, and there is a possibility for the defect to reach all the way down to the interface in these areas. We also noticed an increased presence of this morphology along surface defects, as shown in Figure S3. Moreover, we also observed multiple rotated domains, similar to those observed in Ref [16], forming all along the nanowire.

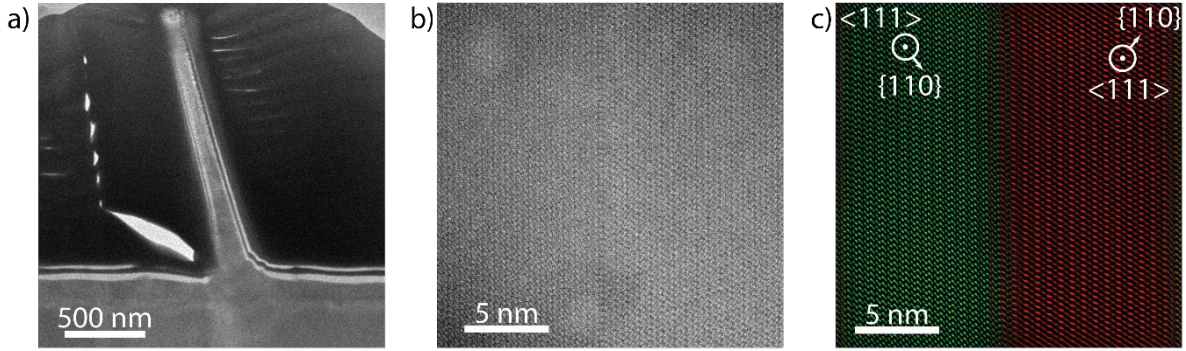

**Figure 5.** (a) Low-magnification view of a cross-section of the new morphology of nanowire. (b) AC HAADF-STEM of the heterotwin running along the axis of the nanowire in (a). (c) colour coded image based on the filtered FFT image of (b), highlighting the different rotations on the different sides of the twin.

### 3.3 Optoelectronic characterisation

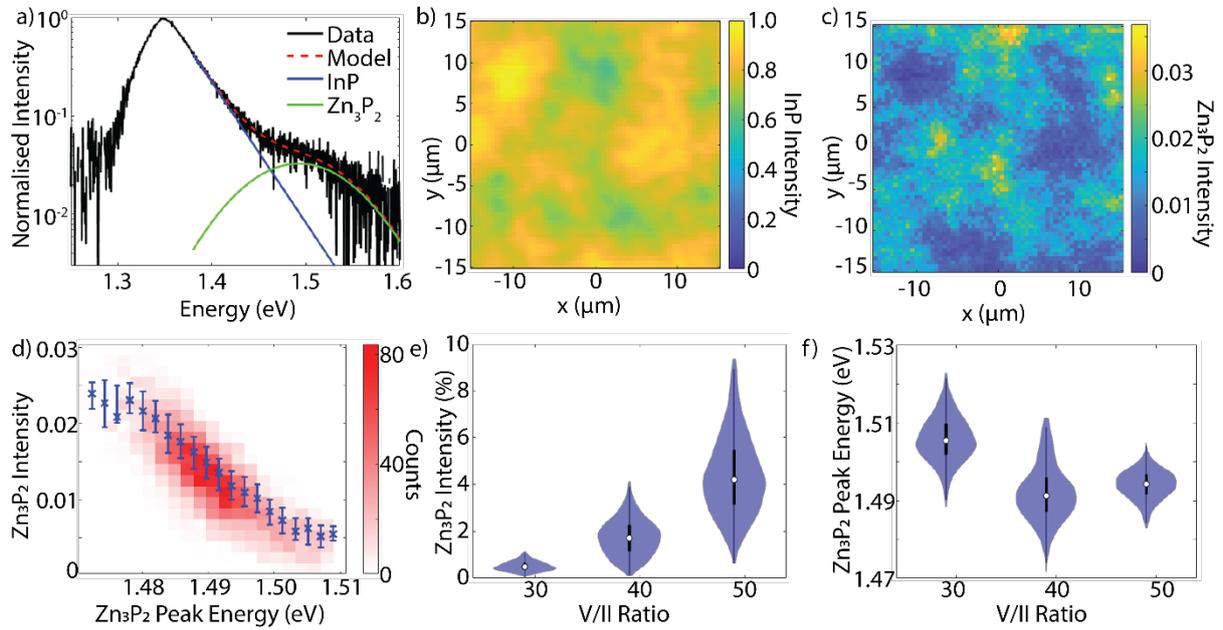

**Figure 6.** Room temperature photoluminescence of $Zn_3P_2$ nanowires on InP substrates. (a) Normalised emission spectrum with emission peaks from the substrate and the nanowires, equation 1 has been fit to the data. (b) Emission intensity map of the InP substrate. (c) Emission intensity of the $Zn_3P_2$ nanowire peak, normalised to the maximum InP emission. (d) 2D histogram showing the correlation between the nanowire peak energy and emission intensity. The blue markers represent the median values in each horizontal bin, with the error bars indicating the interquartile range. The Pearson's linear regression coefficients for this correlation are r = -0.81, p~0. (e-f) Violin plots showing the distribution of the nanowire emission intensity and peak energy for samples grown with different V/II ratios.

To evaluate the optoelectronic properties of the $Zn_3P_2$ nanowires we performed PL mapping of 3 samples grown at 336°C at different V/II ratios. An example PL spectrum from $Zn_3P_2$ nanowires grown with a V/II ratio of 46.3 is shown in Figure 6a. The spectrum contains two features: the dominant peak at 1.35 eV is consistent with band-edge recombination in InP and is therefore attributed to carrier recombination in the InP substrate.[39] On the high energy side of this peak there is a broad shoulder which is 2 orders of magnitude lower in intensity. This shoulder has a peak energy of around 1.5 eV which is comparable to previous studies of $Zn_3P_2$ at room temperature– we therefore attribute this peak to recombination from the $Zn_3P_2$ nanowires.[7,9,15] The comparatively low emission intensity is partially attributed to the reduced volume of $Zn_3P_2$ when compared with the substrate, with an additional contribution from inefficient carrier recombination in the nanowires, potentially due to non-radiative recombination at surface states and planar or point defects related to off-stoichiometric composition.[7,14]

To investigate the spatial dependence of this peak, the spectrum was fit was fit with an empirical model, described by equation 1:

$$I_{PL}(E) = A_{InP} \exp(-k_{InP}(E - E_{InP})) + A_{Zn_3P_2} \exp\left(-\frac{(E - E_{Zn_3P_2})^2}{\sigma^2}\right) \quad (1)$$

where the tail of the InP is represented by an exponential decay, which has an amplitude of $A_{InP}$ at an energy of $E_{InP}$ and a decay constant $k_{InP}$. The higher energy emission is described by a Gaussian with amplitude $A_{Zn_3P_2}$, centred at energy $E_{Zn_3P_2}$ and with width σ. The spatial map of the normalised integrated intensity of the two peaks are shown in Figure 6a and b, respectively. The InP intensity varies by up to 50% over the studied area with correlated regions of 5-10 μm in size. In contrast, the $Zn_3P_2$ intensity changes from zero to 3.5% of the total intensity and is negatively correlated with the InP intensity. This pattern is consistent with a variable nanowire density across the InP substrate: regions of higher nanowire density result in greater photon absorption in the nanowires and thus a higher nanowire emission intensity, along with less absorption in the underlying InP and thus a reduction in the InP emission intensity.

The emission intensity is associated with other optoelectronic properties of the nanowires. Figure 6d shows a strong negative correlation between the emission intensity and the peak photon energy. The peak energy blueshifts by 40 meV across the dataset, and this corresponds to a drop in emission intensity by close to 100%. In $Zn_3P_2$, peak energy shifts have been previously linked to changes in the crystal composition, and these results suggest that this composition is also connected to the carrier recombination efficiency.[7]

To investigate this effect further, an additional two samples were grown with differing V/II ratios. Micro-PL maps were measured from each of these samples and are qualitatively similar in their spatial dependence to those in Figures 6b and c. Despite this, there are stark differences in intensity and emission energy for each of these samples. Figure 6e shows a strong positive relationship between the median $Zn_3P_2$ intensity and the V/II ratio, which increases from 0.5 to 4.2% across the range. SEM image analysis of these samples demonstrate that there is no significant difference in the nanowire density– it is therefore likely that this intensity difference is related to the radiative recombination efficiency. This result implies that, as the V/II ratio is increased during growth, the resultant nanowires have fewer defects and thus the non-radiative recombination rate is reduced in the investigated range.

Figure 6f shows that the V/II ratio also has an impact on the emission energy of the nanowires: the median energy changes from 1.505, to 1.491 and to 1.494 eV, with increasing V/II ratio. Whilst this is a non-monotonic behaviour, the 13 meV shift is statistically significant and suggests a change in the crystal composition and functional properties with V/II ratio. Furthermore, the interquartile range reduces from 8 to 5 meV for the largest V/II ratio studied – suggesting that the crystal uniformity based on the radiative recombination efficiency of the nanowires is improved under these conditions.

## 4. Conclusion

In this study we have demonstrated the steps to grow $Zn_3P_2$ nanowires with high morphological selectivity using In as a catalyst. We found it imperative to load the particles with Zn before providing P to avoid consumption of the catalyst and formation of unwanted phases. Furthermore, we have demonstrated improved morphological control over the $Zn_3P_2$ nanowires by introducing a separate particle deposition step. Using this approach we have mapped out the parameter space to achieve epitaxial growth of $Zn_3P_2$ nanowires on InP (111)A/B substrates using MOCVD and relating it to thermodynamically simulated phase diagrams and experimental limitations, such as $PH_3$ cracking (low T) and Zn re-evaporation (high T).

We performed structural analysis by investigating FIB cross-sections with AC HAADF-STEM, showing that $Zn_3P_2$ nanowires readily grow on either A or B polar InP (111) substrates, however, twins consistently form in the top layers of the InP (111)A substrates. Moreover, we observed misfit dislocations in the thin films growing between nanowires. Moreover, to evaluate the strain due to lattice mismatch we performed GPA. The analysis showed that the nanowire had been able to almost completely accommodate the strain due to lattice mismatch, exhibiting a lattice parameter close to the relaxed bulk lattice parameter from literature. We also investigated a new morphology of nanowires, growing with a rectangular cross section and at an angle to the InP (111) A/B substrate surfaces. These nanowires exhibited a high concentration of defects, such as rotated domains, and heterotwin along the longitudinal axis of the nanowire. The origin of this morphology is suspected to be linked to surface defects in the substrate.

Through PL spectroscopy we investigated the effect of V/II ratio on the functional properties of the $Zn_3P_2$ nanowires. First, for all samples we observed PL from the $Zn_3P_2$ nanowires in the region of 1.5 eV, which is in the range optimal for single-junction solar cells. For the explored V/II range we also observed an increase in PL intensity for samples grown at higher V/II ratio, suggesting that the carrier recombination efficiency in $Zn_3P_2$ is related to the stoichiometry. Furthermore, we observed a decrease of the room temperature bandgap recombination by 13 meV when going from a V/II ratio of 28 to 38, further highlighting the importance of composition in determining the functional properties of $Zn_3P_2$ nanowires.

## 5. Acknowledgements

S. E. S. acknowledges support from the Wenner-Gren Foundation, Åforsk Foundation and Crafoord foundation. The authors also acknowledge support from NanoLund, the Lund Nano Lab (myfab Lund), and Horizon Europe through the Pathfinder project SOLARUP (project number: 101046297). ICN2 acknowledges funding from Generalitat de Catalunya 2021SGR00457. ICN2 is supported by the Severo Ochoa program from Spanish MCIN / AEI (Grant No.: CEX2021-001214-S) and is funded by the CERCA Programme / Generalitat de Catalunya. Part of the present work has been performed in the framework of Universitat Autònoma de Barcelona Materials Science PhD program. Funded by the Horizon Europe call HORIZON-INFRA-2021-SERV-01 project ReMade@ARI under grant agreement number 101058414. Authors acknowledge the use of instrumentation as well as the technical advice provided by the Joint Electron Microscopy Center at ALBA (JEMCA). ICN2 acknowledges funding from Grant IU16-014206 (METCAM-FIB) funded by the European Union through the European Regional Development Fund (ERDF), with the support of the Ministry of Research and Universities, Generalitat de Catalunya. ICN2 is founding member of e-DREAM.[40] P. P. acknowledges support under the UKRI Future Leaders Fellowships scheme (MR/Y03421X/1).

## 6. Notes

The authors declare no conflicts of interest.

# Supplementary information to "Exploring interfaces of epitaxially grown Zn$_3$P$_2$ nanowires and their composition dependent optoelectronic properties"

Simon Escobar Steinvall[1,2], Francesco Salutari[3], Jonas Johansson[2,4], Ishka Das[5], Sebastian Lehmann[2,4], Stephen A. Church[5], M. Chiara Spadaro[3,6], Patrick Parkinson[5], Jordi Arbiol[3,7], Kimberly A. Dick[1,2]

1. Center for Analysis and Synthesis, Lund University, Box 124, 221 00 Lund, Sweden
2. NanoLund, Lund University, Lund, Sweden
3. Catalan Institute of Nanoscience and Nanotechnology (ICN2), CSIC and BIST, 08193 Barcelona, Catalonia, Spain
4. Division of Solid State Physics, Lund University, 221 00 Lund, Sweden
5. Department of Physics and Astronomy and The Photon Science Institute, The University of Manchester, Manchester, M13 9PL United Kingdom
6. Department of Physics and Astronomy "Ettore Majorana", University of Catania and CNR-IMM Via S. Sofia 64, 95123 Catania, Italy
7. ICREA, 08010 Barcelona, Catalonia, Spain

While not explicitly investigated, we did notice a history effect on the growth. At the start of a growth session, we always had a clean chamber and an InP base cover. With subsequent growth we did see the formation of a Zn-P cover in the chamber. There was an influence on the growth rate, mainly between the first growth and second growth where it increased, after which it quickly stabilises. However, the main influence was on the In nanoparticle formation. After some experiments we started to observe a decrease in the nanowire density, and eventually the nanowire formation was supressed completely. In short, the Zn-P cover interferes with the In nanoparticle deposition. We could return back to initial conditions by re-depositing our InP base cover, resetting any history effects.

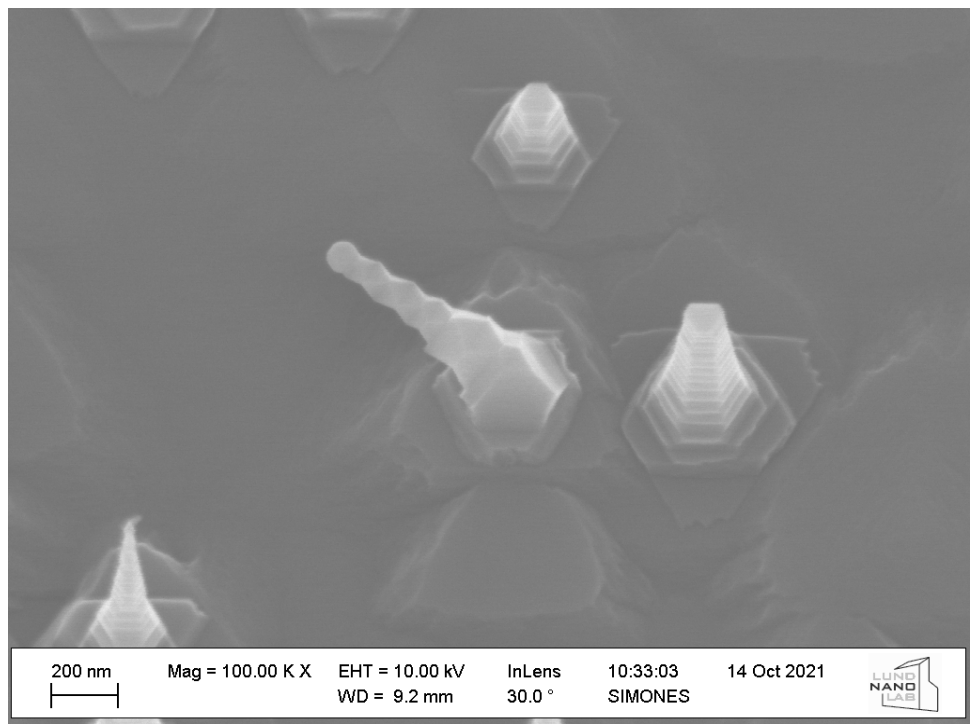

**Figure S1.** High magnification SEM image of nanowire at high V/II without catalyst.

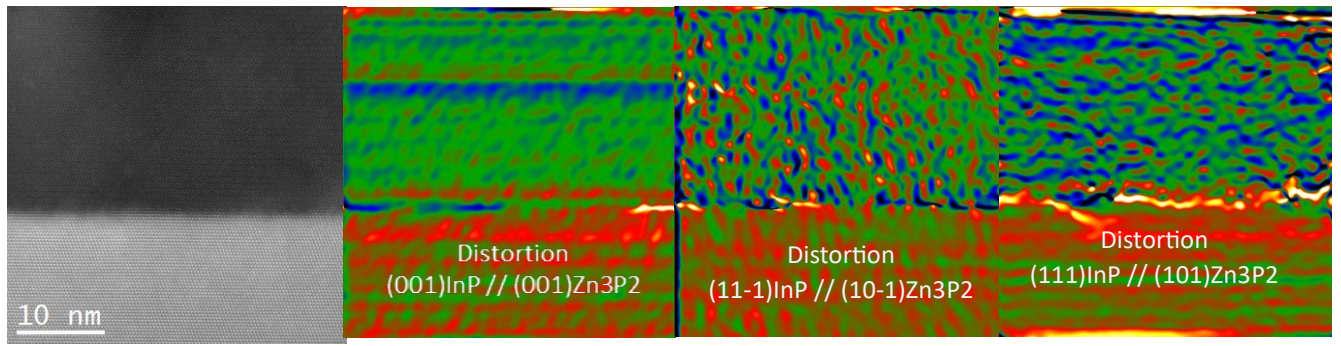

**Figure S2.** AC HAADF STEM image and corresponding GPA map of interface between InP (111)B and $Zn_3P_2$ nanowires showing no regular misfit dislocations. The blue horizontal line observed in the dilatation maps for the (001) planes is due to a STEM scanning artifact, and not a defect in the material.

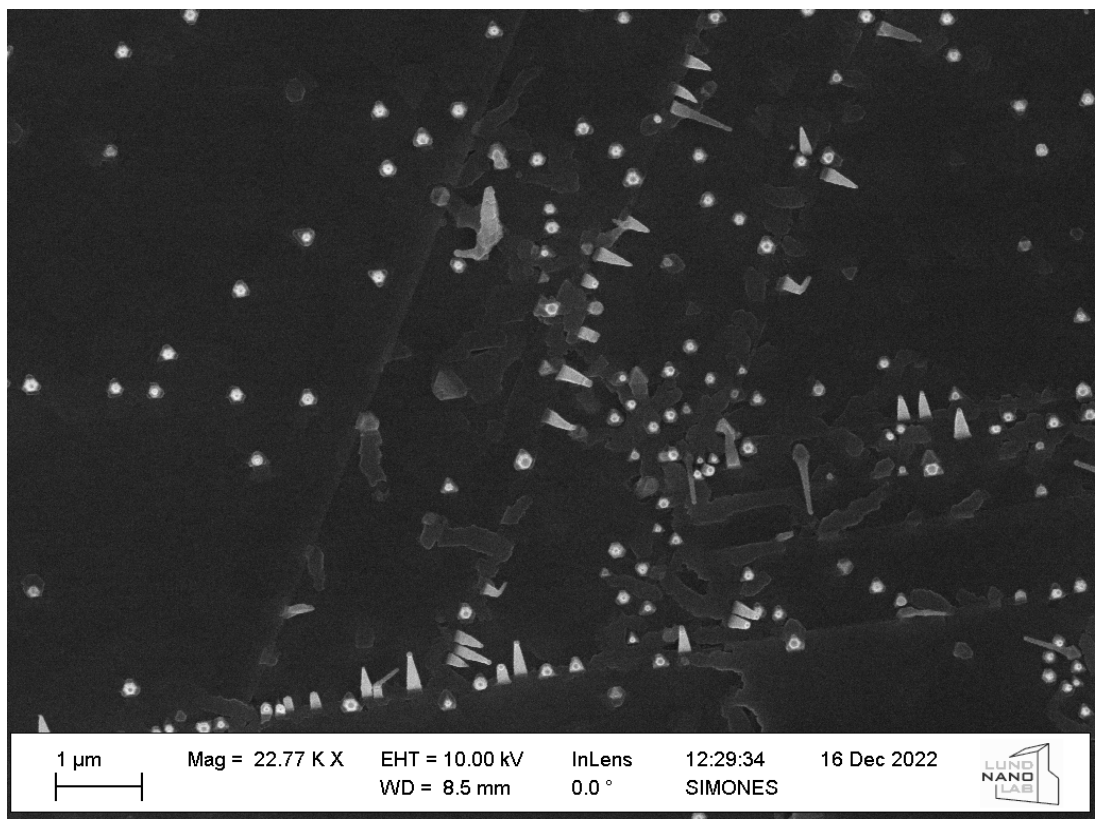

**Figure S3.** SEM of tilted nanowires growing with higher density from a surface defect.